\documentclass[applsci,article,accept,moreauthors,pdftex]{mdpi} 
\usepackage{color} 
\usepackage{graphics}
\usepackage{graphicx}
\usepackage{amssymb}
\usepackage{amsfonts}
\usepackage{caption}
\usepackage{hyperref}
\usepackage{caption}
\usepackage[labelformat=simple]{subcaption}

\firstpage{1} 
\pubvolume{xx}
\issuenum{1}
\articlenumber{5}
\pubyear{2019}
\copyrightyear{2019}
\history{Received: 4 April 2019; Accepted: 8 June 2019; Published: date}
\updates{yes} 

\Title{A Self-Consistent Quantum Field Theory for Random~Lasing}

\Author{ Andreas Lubatsch $^{1,2,\dagger}$ and Regine Frank $^{1,3,}$*$^{,\dagger}$}

\AuthorNames{Andreas Lubatsch  and Regine Frank}

\address{%
$^{1}$ \quad Physikalisches Institut, Rheinische Friedrich-Wilhelms
  Universit\"at Bonn, Wegelerstr. 8, 53115 Bonn, Germany; lubatsch@th.physik.uni-bonn.de\\
$^{2}$ \quad Georg-Simon-Ohm University of Applied Sciences, Ke{\ss}lerplatz
  12, 90489 N\"urnberg, Germany\\

$^{3}$ \quad Serin Physics Laboratory, Department of Physics and Astronomy,
Rutgers University, 136 Frelinghuysen Road, Piscataway, NJ 08854-8019, USA; regine.frank@rutgers.edu}

\corres{Correspondence: regine.frank@googlemail.com}

\firstnote{These authors contributed equally to this work.}

\abstract{The spatial formation of coherent random laser modes in strongly
  scattering disordered random media is a central feature in the understanding
  of the physics of random lasers. We derive a quantum field theoretical
  method for random lasing in disordered samples of complex amplifying Mie
  resonators which is able to provide self-consistently and free of any fit
  parameter the full set of transport characteristics at and above the laser
  phase transition. The coherence length and the 
  correlation volume respectively is derived as an experimentally measurable
  scale of the phase
  transition at the laser threshold. We find that
the process of stimulated emission in extended disordered arrangements of
active Mie resonators is
ultimately connected to time-reversal symmetric multiple
scattering in the sense of photonic transport while the diffusion coefficient
is finite. A power law is found for the random laser mode diameters  in stationary state with increasing pump
intensity.}


\begin{document}


\section{Introduction}
\label{INTRO}

\begin{figure}[H]
  \hspace*{0.0cm}{\scalebox{0.78}{\includegraphics{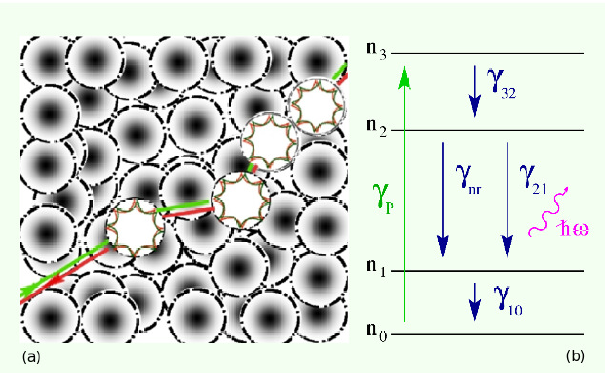}}}\quad{\scalebox{0.36}{\includegraphics{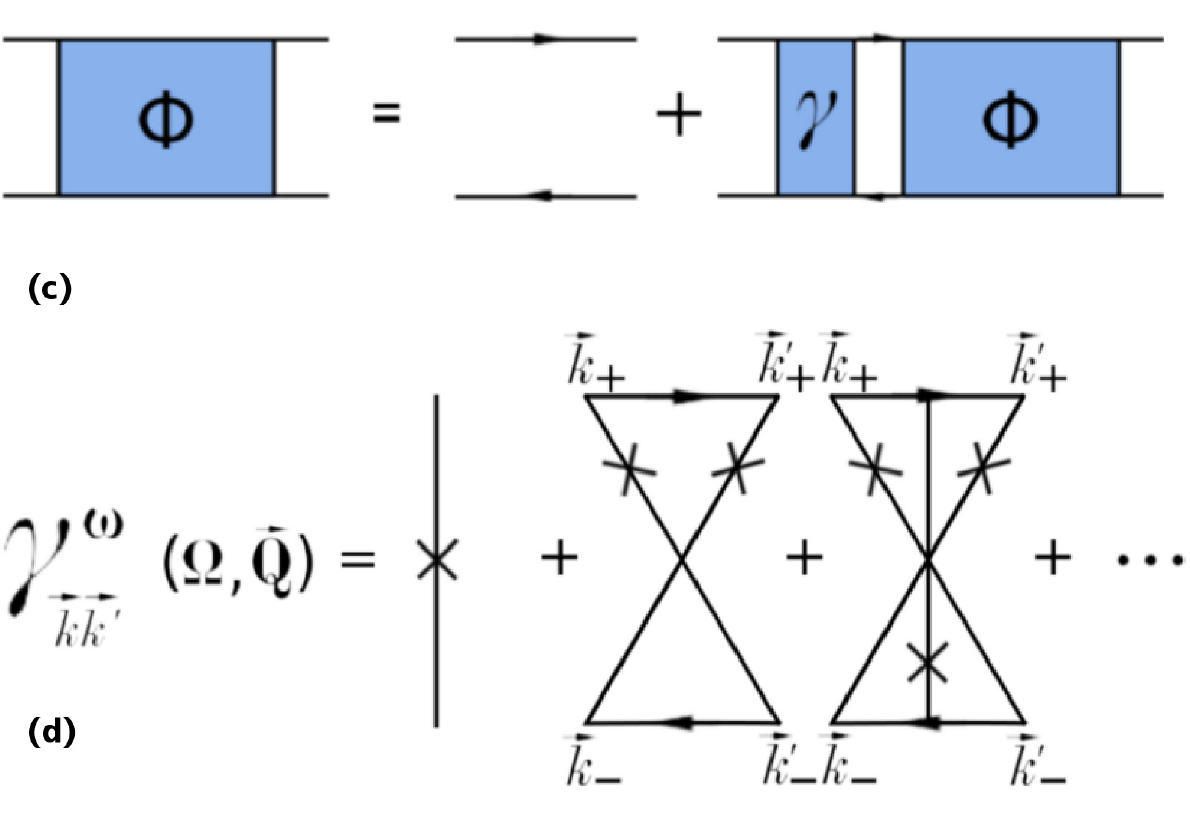}}}
\vspace*{0.1cm}       
\caption{(\textbf{a}) Sketch of a disordered semi-conductor random laser slab,
  that is open in $z-$direction. Photonic transport processes (red) and their
  time reversal (green) interfere while they feature multiple scattering
  processes with complex active Mie resonators. The~Mie resonance~\cite{Mie}
  is a so called whispering gallery resonance at the inner surface of the independent Mie scatterer. We use for this work
  monodisperse Mie spheres. (\textbf{b}) 4-level laser rate equation scheme. Straight lines
  represent the electronic processes, the~green line represents the
  excitation due to the pump field, $\gamma_{32}$ represents the decay from
  level $n_3$ to the upper laser level $n_2$, $\gamma_{nr}$ represents
  nonradiative decay and as its
  competitor $\gamma_{21}$ leads to spontaneous and to stimulated emission processes, which
  eventually yields inversion and coherent laser radiation $\hbar\omega$. (\textbf{c}) Schematic representation of the
  {\em Bethe-Salpeter equation}, compare Equation~(\ref{BETHESALPETER}). $\Phi$ is
  the intensity field in the sample, $\gamma_{\vec k\vec k'}^{\omega}$ is the
  irreducible vertex. (\textbf{d})~The irreducible vertex includes all orders of maximally 
crossed diagrams (Cooperon) which represent all quantum-coherent interference
contributions due to multiple scattering in presence of disorder.}\label{fig. 1}
\end{figure}

The research for random lasers is an emerging research field
~\cite{Cao99,CAOPL,Mujumdar3,Wiersma_NATPHYS, Noginov,Vardeny, Mujumdar1, Mujumdar2, Niyuki, Chen1, Krauss},
which recently has been extended to highly flexible~\cite{Lau} and unconventional
materials and setups~\cite{Popov1,Popov2}.  If~these systems can be operated
spectrally and spatially well controlled they feature the future as
large area coherent light sources ahead of all state of the art
LEDs. Theoretically many different models from statistical
physics~\cite{HackenbroichL,Mujumdar4,Hackenbroich,Mujumdar6,Lepri1,Henneberger,Lepri2,Gaio},
classical field theoretical methods~\cite{Soukoulis1,Soukoulis2,Tuereci,Stone,Conti}, quantum dynamical
~\cite{Versteegh1,Versteegh2,Lodahl2D} and quantum field theoretical approaches
~\cite{Frank06,JOA09,ANN09, Frank11} for embedded disordered
  ensembles of laser active scatterers ~\cite{NJP14,SREP15} are investigated. The~random laser setup consists of a multiple scattering medium which can be
also passive Mie scatterers infiltrated by laser active dye
~\cite{Gottardo} and dye infiltrated disordered waveguides~\cite{Consoli}. Random lasers are operated in absence of any
external feedback or mirror system, thus it is of principal importance that a high
contrast of the refractive index between scatterer and background is given in
order to enhance multiple scattering and thus photonic transport and as a
result of the accumulation of a high number
of photons in the sample. A~scheme of a possible setup for monodisperse
solid state Mie scatterers is found in Figure~\ref{fig. 1}a. Monodisperse Mie
spheres can become in certain configurations extremely sensible systems,
especially when the scatterers are large compared to the transport
wavelength. In~this regime the Mie spheres support the occurrence
  of intensity fluctuations. We~chose a
system of monodisperse scatterers for this article in order to check
systematically whether the signature of the independent single Mie  scatterer in the transport characteristics of the non-pumped and non-inverted microscopic system will
  persist in the stationary state of lasing. We develop a quantum field theoretical approach of photonic transport based on
the {\em Bethe-Salpeter equation}, see Figure~\ref{fig. 1}c, that incorporates all orders of interference
effects by means of the Cooperon, the~maximally crossed Feynman diagram
~\cite{VW}, see
Figure~\ref{fig. 1}d. Energy conservation laws are
  implemented  by means of a
generalized Ward identity~\cite{Lubatsch05}. We couple this framework to the microscopic laser rate
equations for quantum cascades, see  Figure~\ref{fig. 1}b, that ensure
particle conservation on the microscopic~level.


For three dimensional random lasers of finite and possibly large extent
theoretical frameworks often come along with tremendous computational
efforts. Our approach is one theoretical possibility to deal with dense
ensembles of strongly scattering resonators efficiently, either in the
case of independent scatterers or in the case of photonic glasses. We derive
in the following a self-consistent frame which provides systematic results
free of any fit parameter that are directly measurable in random laser
experiments.

\section{Quantum-Field Theory for Photonic~Transport}
\label{QFT}

We develop in this article a self-consistent quantum field theory for random
lasing of ensembles of active Mie resonators. The~results are spatial
characteristics of the phase transition to random lasing, they classify the
threshold to stationary state lasing. The~theory includes multiple
scattering effects of photons in complex random arrangements which may be
classified in incoherent contributions that are expressed in quantum field
theory, e.g.,~as ladder diagrams, the~diffuson, and~coherent contributions, that
are expressed in terms of maximally crossed diagrams, the~Cooperon. Methodologically the Bethe-Salpeter equation, see
Figure~\ref{fig. 1}c, is a most reasonable self-consistent frame, where the so
called memory kernel $M$ as part of the irreducible vertex $\gamma$ bares the expansion in orders of maximally
crossed diagrams, Figure~\ref{fig. 1}d. All light-matter
interactions are represented within the irreducible vertex, e.g.,~the interaction of the
electromagnetic field and its time reversal as well as the light intensity
with a complex Mie resonator as a scatterer~\cite{Lubatsch05}. Samples of a
large number of active Mie resonators can be considered,
which may act as independent scatterers or as {\it glasses}, i.e.,~as
an accumulation of agglomerates with an intrinsic correlation. In~what follows we derive this theory for the case of monodisperse independent
active Mie scatterers, where pronounced dips were found theoretically and
experimentally in the results for the transport characteristics, e.g.,~the
scattering mean-free path $l_s$ \cite{Frank11}.

\subsection{Nonlinear~Response}

The underlying electrodynamics for transport in non-linear media is described basically by the wave equation for Kerr-Media
\begin{eqnarray}
\Delta \vec E - \frac{\epsilon}{c^2}\frac{\partial^2 \vec E}{\partial t^2} =
\frac{1}{c^2\epsilon_0}\frac{\partial^2\vec P}{\partial t^2}.
\label{NL}
\end{eqnarray}
where the polarizability $\vec P(\vec E)$ may be decomposed in linear and
non-linear part $\vec P(\vec E) = \epsilon_0(\epsilon -1)\vec E + \vec P_{NL}$.
The electric displacement inside a medium is written as $\vec D(\vec E) =
\epsilon_0 \epsilon \vec E + \vec P_{NL}$. Kerr media are $\chi^{(2)}$ or higher order processes in their dependency to the electrical field $E$.
\begin{eqnarray}
\vec P = \epsilon_0 \chi^{(1)} \vec E + \epsilon_0 \chi^{(2)} \vec E\cdot\vec
E +...
\end{eqnarray}

\subsection{Bethe-Salpeter Equation for Photonic Transport in Samples of Disordered Active Mie~Resonators}

We consider the propagating photonic intensity, see
Figure~\ref{fig. 1}c, as~the field field correlator $\Phi$ which is derived
selfconsistently by the {\em Bethe-Salpeter equation}. The~propagating
electromagnetic wave in disordered random media is described by the
single-particle Green's function Equation~(\ref{GreenSP}) that solves the non-linear
wave equation, Equation~(\ref{NL}).
\begin{eqnarray} 
G^{\omega}_{\vec k} = \frac{1}{\epsilon_b(\omega/c)^2 - |\vec k|^2
  -\Sigma^{\omega}_{\vec k}}
\label{GreenSP}
\end{eqnarray} 

 The scatterers are embedded in a background matrix, with~the
   dielectric function $\epsilon_b$, e.g.,~in air. The~granularity of the random medium is implemented in the form of a spatially dependent potential of the permittivity
$\epsilon_s(\vec r)$. As~a result of our theoretical framework we
derive the laser gain self-consistently and space-resolved. The~order of non-linearity is a matter of the numerical
self-consistency as the independent complex semiconductor scatterers 
are well described by the T-matrix~\cite{Lubatsch05} for spherical
Mie resonators of matter with a self-consistent complex refractive
index.

We use a model of independent Mie scatterers, Figure~\ref{fig. 1}b, here, in order
to derive the
self-energy contribution $\Sigma^\omega_{\vec k}$ for extended photo-active
particles in Equation~(\ref{GreenSP}). The~Mie scattering coefficients of n-th order are known from the
literature~\cite{Mie} to have the following form
\begin{eqnarray} 
a_n &=&
\frac{
    m\Psi_n(my)\Psi_n^{\prime}(y)
     - \Psi_n(y)\Psi_n^{\prime}(my)
}
{
      m\Psi_n(my)\xi_n^{\prime}(y)
     -   \xi_n(y)\Psi_n^{\prime}(my)}\\\nonumber
\\\nonumber	
b_n &=&
\frac{
    \Psi_n(my)\Psi_n^{\prime}(y)
     -m\Psi_n(y)\Psi_n^{\prime}(my)
}
{
    \Psi_n(my)\xi_n^{\prime}(y)
     -m\xi_n(y)\Psi_n^{\prime}(my).	
}\\\nonumber
\end{eqnarray}

In this notation $m = \frac{n_s}{n_b}$ is the relative refractive index
between the scatterer and the background matrix, $y=\frac{2\pi n_sr_S}{\lambda}$ is the size
parameter depending on the scatterers' radius $r_S$ as well as on the transport
wavelength $\lambda$. $\Psi_n$, $\xi_n$ are Riccati-Bessel functions.
The refractive index $n_s(\Phi)$ obtains  a higher-order
non-linearity as consequence of the self-consistency with respect to $\Phi$,
${\rm Im} \epsilon_s(\Phi)$. The~independent scatterer approach works very
well in dense random media of filling of 35\%--55\%. Positional dependent
enhancement effects due to the high filling fraction are effectively mapped on the dynamics of $n_s(\Phi)$ of
the single scatterer in non-linear response, and~we use the exact Mie scattering
solution.

The Green's function, Equation~(\ref{GreenSP}), for~$\Psi_{\omega}$, solves the wave
equation Equation~(\ref{NL}) and builds up the intensity field in the sample in the form of the
field correlator $\Phi^{\omega}_{ \vec{k} \,\, \vec{k}' } (\vec{Q},\Omega)$. The~four point~correlator
\begin{equation} 
\Phi^{\omega}_{ \vec{k} \,\, \vec{k}' } (\vec{Q},\Omega) = \langle
\hat{G}^R_{\vec{k}_+\,\,\vec{k}'_+ } (\omega)
\hat{G}^A_{\vec{k}_-\,\,\vec{k}'_- } (\omega)\rangle
\label{FourP}
\end{equation} 

\noindent is thus expressed in terms of retarded and advanced Green's function $\hat{G}^{R/A}$ denoting the field and its time-reversal. 
We utilize the transformation of coordinates in center-of-motion $(\vec{Q},\Omega)$ and relative $(\vec{k},\omega)$
momenta and frequencies~\cite{Frank11} $\vec{k}_{\pm} = \vec{k} \pm \vec{Q}/2 $ and $\omega_{\pm} = \omega \pm
\Omega/2$. Considering the slab geometry to be extended within the 
$(x,y)$-plane and finite in $z$-direction, the~full Fourier transform as
indicated for infinite samples~\cite{Frank06} is replaced by a partial Fourier
transform following the argument of the separation of the scales for the field
$\Psi$ and the intensity $\Phi$. $\Psi$ is characterized
by the wavelength $\lambda$, whereas the change of the light intensity $\Phi$
is characterized by the transport mean-free path $\xi$, and the
correlation length in random lasers respectively, which is one of the
central characteristics that we derive in due
course of the paper. In~$(x,y)$-plane the standard Fourier transform is used, in~the limited $z$-direction we Fourier-transform the relative
coordinate but the center-of-motion coordinate $Z$ remains in real
space. The~separation of scales allows for the physical incorporation of loss
at the samples boundaries at the level of the transport theory. The~ break of
the T-symmetry, the~break of the time reversal invariance of the multiple
scattering processes, is~incorporated in a macrocanonical sense
~\cite{Onsager} on the level of the integration of the real space coordinate $Z$.  
We apply these arguments to the equation of motion for the intensity
correlation,  known as the {\em Bethe-Salpeter equation}, see Figure~\ref{fig. 1}c,
\begin{equation}
\Phi_{\epsilon\epsilon} = G^RG^A[1 + \int\frac{d^3k}{(2\pi)^3}
\gamma\Phi_{\epsilon\epsilon}]
\label{BETHESALPETER}
\end{equation}

\noindent and we obtain the Boltzmann or kinetic equation for
transport Equation~(\ref{bethe-salpeter}). We introduce here the following
abbreviations, consistent with previous work, $\Delta\Sigma =
\Sigma^A(\omega_-)-\Sigma^R(\omega_+)$, $\Box \Sigma = \Sigma^A(\omega_-) + \Sigma^R(\omega_+)$, and~equivalent expressions for
$\Delta G^{\omega_-}_{\vec{k}-}$ and $\Box G^{\omega_-}_{\vec{k}-}$.
The term $\gamma^{\omega}_{ \vec{k} \,\, \vec{k}\,''} (Z,Z\,'',\vec{Q}_{||},\Omega )$ 
represents the irreducible vertex, which is physically interpreted here as the
coherent light matter interaction in disordered granular non-linear
systems. The~Ward identity is derived in the generalized form for
  the scattering of photons in {\it non-conserving} media.  Absorption or gain
  yield an additional contribution, and~a form of the Ward-Takahashi identity for photons in complex matter~\cite{Lubatsch05,WARD,TAKAHASHI} is
  derived. Effectively the additional contribution is not negligible and thus presents in the
theoretical results of the transport characteristics of the self-consistent
framework. It also renormalizes the energy
transport velocity~\cite{Lubatsch05,Frank06,Frank11}. The~solution to
Equation~(\ref{BETHESALPETER}), is derived as the energy density response
$\Phi_{\epsilon\epsilon}$ in the form of a diffusion pole, Equation~(\ref{pole})
\begin{eqnarray}
\label{pole}
\Phi_{\epsilon\epsilon}
=
\frac{ N_{\omega} }
{
\Omega + i D Q^2_{||} + iD \xi^{-2}
}.
\end{eqnarray}
\begin{eqnarray} 
\label{bethe-salpeter}
\big[2{\rm Re\,}\left(\epsilon_b\right) \omega\Omega
\,\,&-&\,\,2\vec{k}_{||}\cdot \vec{Q}_{||}
+
2ik_z\partial_Z
\,\,+\,\,\Delta \Sigma 
\,\,-\,\,
\Delta \epsilon_b \omega^2
\big]\,\Phi^{\omega}_{ \vec{k} \,\, \vec{k}\,' } ( Z,Z',\vec{Q}_{||},\Omega)
\\
&=& \Delta G \delta ( \vec{k}-\vec{k}\,'\, ) \,\, +\,\, \sum_{Z\,''}\!
\Delta G \!\!\int \!\!\frac{ {\rm d}^3k\,''}{(2\pi)^3} 
\gamma^{\omega}_{ \vec{k} \,\, \vec{k}\,''} ( Z,Z\,'',\vec{Q}_{||},\Omega )
 \Phi^{\omega}_{ \vec{k}\,'' \,\, \vec{k}\,' } (
 Z\,'',Z\,',\vec{Q}_{||},\Omega )\nonumber\\\nonumber
\end{eqnarray} 
\begin{equation}
M(\Omega)
=
\frac
{
1
}
{
\int\frac{{\rm d}^3k}{(2\pi)^3} 
(2\vec{k}\cdot\hat{Q})
(\Delta G)^2
}
\int\frac{{\rm d}^3k}{(2\pi)^3} 
\int\frac{{\rm d}^3k'}{(2\pi)^3}
\,(2\vec{k}\cdot\hat{Q}) \Delta G_k
\gamma_{kk'}
(2\vec{k}\,'\cdot\hat{Q}) (\Delta G_{k'})^2
\label{memory}
\end{equation}

We define the correlation length $\xi$ in dependency to the energy density
$\Phi_{\epsilon\epsilon}$ itself
\begin{eqnarray} 
\label{TT_aux_230} 
\frac{1}{\xi^2(Z)} 
= 
\frac {N_{\omega} }{ D(\Omega=0; Z) \Phi_{\epsilon\epsilon}(Q_{||}=0, Z;\Omega=0)}.
\end{eqnarray}

The full diffusion coefficient $D(\Omega=0; Z)$ is derived as
\begin{eqnarray} 
D(\Omega) = D_0^{tot} -\tau^2D(\Omega)M(\Omega)
\label{Diff1}
\end{eqnarray}
\noindent where the bare diffusion $D_0 = \frac{2v_Ec_p}{\omega \Delta G}
\int\frac{d^3k}{(2\pi)^3} (\vec{k}\cdot\hat{Q})^2\Delta G$ 
is complemented by the contributions originating from the active medium as the
scatterers $D_s = \frac 1 8 r_{\epsilon}A_{\epsilon}\tau^2  \tilde{D}_0$ or
the background $D_b  = \frac 1 4 (\omega\tau)^2 \Delta\epsilon_b
\tilde{D}_0$. Thus the diffusion coefficient without memory effects reads
$D_0^{tot} = D_0+D_s+D_b $. The~term $-\tau^2D(\Omega)M(\Omega)$ contains
the memory kernel $M(\Omega)$, Equation~(\ref{memory}), including the Cooperon
contribution and consequentially all interferences. $\tilde{D}_0$ equals $D_0$ where the imaginary part $\Delta G$ is replaced by the
real part $\Box G$. The~renormalized density of states $N_{\omega}$ is derived
as follows
\begin{eqnarray}
N_{\omega}     = \frac{\omega^2 \Delta G_0 (\vec{Q}, \Omega)} 
     {c_p^2 g_{\omega}^{(1)}[1+ \Delta (\omega)]}.
\label{LDOS}
\end{eqnarray}

We further use the notation and  the abbreviations of~\cite{Frank11}. $\Delta(\omega) = B_{\epsilon}A_{\epsilon} +i
r_{\epsilon}\partial_{\Omega}A_{\epsilon}(\Omega)$ with $A_{\epsilon} = 2
[u _{\epsilon} {\rm Re\,} G_o +  {\rm Re\,} \Sigma_o]$ and $B_{\epsilon}  = \frac{({\rm Re\,}\Delta\epsilon)^2+({\rm Im\,}\Delta\epsilon)^2}
      {2\omega^2({\rm Re\,}\Delta\epsilon)^2}$, $\partial_{\Omega}$ is the
      differential resulting from the expansion in $\vec Q$ and $\Omega$,
      where the momentum integrated form of $\Delta G$ is $\Delta G_0 = \int \frac{ {\rm d}^3
        k}{(2\pi)^3} \Delta G_{p}^{\omega}(Q,\Omega,Z)$ and $\Delta
      \epsilon = \epsilon_s  - \epsilon_b$, $u _{\epsilon}  =\frac {{\rm Im\,} (\Delta\epsilon \Sigma^{\omega})}
   {{\rm Im\,} (\Delta\epsilon G_0^{\omega}) }$ and the abbreviations $r_{\epsilon}   = {{\rm
       Im\,} \Delta\epsilon}/{{\rm Re\,} \Delta\epsilon}$ and
   $g_{\omega}^{(1)} = \frac {4\omega}{c^2}{\rm Re\,} \epsilon_b$, as~well as $g_{\omega}^{(0)} = \frac {2\omega}{c^2}{\rm Im\,}
   \epsilon_b$ are~introduced. 

The energy transport
   velocity is derived as $v_E=\frac{c^2}{c_p{\rm
       Re}\epsilon_b}\frac{1}{1+\Delta(\omega)}$, whereas $c_P={\rm Re}\frac{c}{\sqrt{\epsilon_b-\Sigma^{\omega}_{0}\frac{c^2}{\omega^2}}}$ equals the
   phase velocity, each one is self-consistently derived. A~detailed discussion
   of the time scale $\tau$ is found in reference~\cite{Frank11}. By~solving the renormalized diffusion equation
Equation~(\ref{diff})
\begin{eqnarray}
i D \xi^{-2}= - i D
  {\chi_{d}^{-2}}-{c_{1}\Big(\partial^{2}_{Y}\Phi_{\epsilon\epsilon}(Q,\Omega)\Big)+
    c_{2}}.
\label{diff}
\end{eqnarray}
coupled with the energy density to the 4-level laser rate equations, see Section~\ref{sec:LRG}, the~coefficients $c_1$ and $c_2$ are self-consistently dereived, and~we arrive the
spatial distribution of energy density
\begin{eqnarray}
-\frac{\partial^2}{\partial Y^2}  \Phi_{\epsilon\epsilon}
=
 \frac{1}{D}\frac{D}{- \chi_d^2}
\Phi_{\epsilon\epsilon}
+ {\rm ASE}.
\label{SE}
\end{eqnarray}

The term for nonlinear self-consistent microscopic random laser gain $\gamma_{21}  n_2$, see Section~\ref{sec:LRG},
incorporates the influence of the boundary renormalized length scale
$\chi_d$; $\gamma_{21}$ is the transition rate from laser level 2 to level 1,
$n_2$ is the electronic occupation number of level 2. This yields the inversion condition
\begin{eqnarray}
\frac{D}{- \chi_d^2} = \gamma_{21}  n_2 
\label{n2}
\end{eqnarray}
in stationary state. $\chi_d$ is the length scale implicated by dissipation in
the bulk alone. The~modification of the boundary, Equation~(\ref{diff}), specifies the relation
between lasing emission and amplified spontaneous contributions (ASE), Equation~(\ref{SE}).

\subsection{Coupling to the Four Level Laser Rate Equations for Quantum~Cascades}
\label{sec:LRG}

The lasing dynamics in our theoretical framework are included by a four-level
laser rate equation system, which is well known from quantum cascade lasers
~\cite{Faist,NJP14,SREP15}

\begin{eqnarray} 
\frac{\partial N_3}{\partial t}  
&=&  
\frac{N_0}{\tau_{P}}  - \frac{ N_3}{\tau_{32}} \\ 
\frac{\partial N_2}{\partial t}  
&=&   
\frac{N_3}{\tau_{32}}  -   \left(\frac{1}{\tau_{21}} 
+ \frac{1}{\tau_{nr}}\right)N_2 -   
\frac{\left( N_2 -N_1\right)}{\tau_{21}} n_{ph} \\ 
\frac{\partial N_1}{\partial t}  
&=&     
\left(\frac{1}{\tau_{21}}+ \frac{1}{\tau_{nr}}\right)N_2   \label{N2}
+ \frac{\left( N_2 -N_1\right)}{\tau_{21}} n_{ph}  
-   \frac{ N_1}{\tau_{10}} \label{third_LRG} \\ 
\frac{\partial N_0}{\partial t}  
&=&    
\frac{N_1}{\tau_{10}} - \frac{N_0}{\tau_{P}} \\ 
N_{tot} &=& N_0 + N_1 + N_2 + N_3, 
\end{eqnarray}

All transition times $\tau_i$ here are given as the inverse of the transition
rates $\gamma_i$, see for details Figure~\ref{fig. 1}b,
$\gamma\,=\,1/\tau_i$. The~numbers represent the laser
levels. $\gamma_{21}\,=\,1/\tau_{21}$ represents the transition from upper to
lower laser level, $\gamma_{P}$ is the pump rate, $\gamma_{nr}$ represents nonradiative
decay. While~we included nonradiative decay processes in previous work~\cite{SREP15} for
polydisperse ZnO powders, they are neglected in this work here. The~numbers $N_i$ represent the level
occupation and their resummation $N_{tot}$ ensures energy as well as particle
conservation. The~number $n_{ph}$ represents the photon numbers
due to spontaneous emission and due to lasing. The~coupling to the transport
theory is given by the stimulated decay procedures, i.e.,~by
Equation (\ref{N2}), which is connected to Equation (\ref{n2}). In~due course of this
article the transition rate $\gamma_{21}$ serves as the measure of the pump strength $P$.

\section{Results and~Discussion}
\unskip

\subsection{The Coherence Volume of D~=~3 Dimensional Random~Lasers}
\label{sec:NUMSOL}

We find a full set of self-consistent scales, e.g.,~the diffusion
coefficient $D$, the~scattering mean-free path $l_s$, the~transport mean-free path
or coherence length $\xi$, the~energy transport velocity $v_E$, the~phase velocity $c_p$, the~energy and intensity dependent LDOS $N_\omega$,  as~well as the laser thresholds and the random laser gain in
stationary state from
the solution of the Bethe-Salpeter equation and the
intensity correlator including interference effects
$\Phi_{\epsilon\epsilon}$. 

\begin{figure}[H]
 \hspace*{0cm}\scalebox{0.15}{\includegraphics{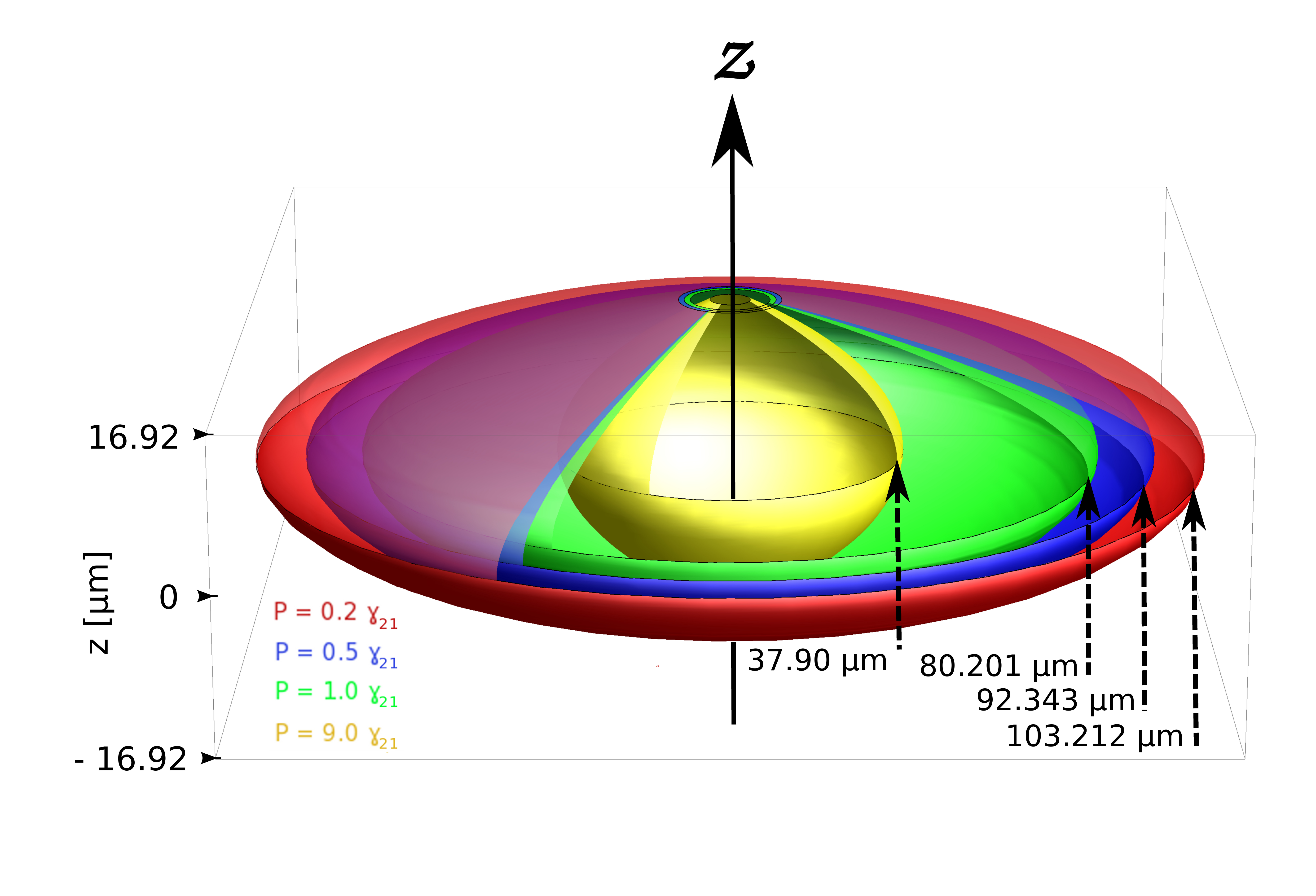}}
\vspace*{0.3cm}
\caption{Coherence volume as the solution from Equation~(\ref{TT_aux_230}) within a D = 3 dimensional random laser
  slab of monodisperse Mie scatterers. The~scatterers' radius is
  $r_{scat}=423.0$ nm, the~particles refractive index is $n\,=\,2.4$, the~real
  part of permittivity is
  ${\rm Re}\,\epsilon_{scat}\,=\, 5.76$, which is comparable to diamond (C). The~transport wavelength for this result is
  chosen as 
  $\lambda\,=\,385$ nm, the~samples filling fraction is chosen as 50$\%$
  volume filling. The~sample is
  of finite size in the z-direction, its extension in z-direction is
  $d\,=\,80\,r_{scat}\,=\,33.84\, \mu m$. The~strongly scattering medium is thus
  open, and~the translation invariance is
  broken in the z-direction, whereas the sample is infinite in the x-y
  plane. We show results for the excitation rates of  $P\,=\,0.2\,\gamma_{21}$ (red), $0.5\,\gamma_{21}$
  (blue), $1.0\,\gamma_{21}$ (green), $9.0\,\gamma_{21}$ (yellow) in units of the stimulated emission
  rate $[\gamma_{21}]$, see Figure \ref{fig. 1}b. The~stationary state
  coherence volume for each value of the excitation power assumes in
  stationary state the form of a three dimensional extended disk, which~is
  symmetrical in plane of the slab. The~radius of the disk for each excitation
  power is noted in the figure. The~disk is a spatial measure
  for the random laser threshold of the specific random laser arrangement. At~the samples boundaries, the~coherence length assumes a finite value which
  depends on the pump strength. It shall be noted that in cases for
  $P\,<\,1.0\,\gamma_{21}$ the external excitation strength is smaller than the
  threshold value for bulk matter. Thus in this regime the multiple scattering
  procedures between Mie resonators are the dominating physical concept for
  the lasing transition of the random scatterer~arrangement.}
\label{vol}
\end{figure}

As the main solution from the generalized diffusion equation, Equation~(\ref{diff}), we
derive the coherence length $\xi$ in a 3-dimensional setup with one finite
dimension $z$. The~finite system is characterized by lossy boundaries,
otherwise we neglect losses for our considerations here in order to derive a
systematic presentation of the random lasing phase transition. The~translation
invariance is consequentially broken in the $z$ direction. We present in
Figure~\ref{vol} results for one specific set of a possible random laser sample under three different
excitation regimes with respect to the microscopic bulk matter properties of
the scatterer's material,  sub-threshold excitation with respect to bulk,
excitation at the bulk threshold and excitation above the bulk~threshold.

We show results for an arrangement of diamond nano resonators of refractive
index $n\,=\,2.4$ with the radius of the single scatterer
$r_{scat}\,=\,423.0\,$ nm in a slab geometry of $ 50\,\%$ filling and the extent of
$80\,r_{scat}\,=\,33.84\, \mu m$ in z-direction whereas it is infinite in-plane. The~sample is equally lossy on either of the boundaries. We assume spatially
homogeneous pumping of the sample and we show the results for the excitation
strengths $P\,=\,0.2\,\gamma_{21}$, $P\,=\,0.5\,\gamma_{21}$, $P\,=\,1.0\,\gamma_{21}$
and $P\,=\,9.0\,\gamma_{21}$, where $\gamma_{21}$ is the rate for stimulated
emission at the lasing transition of the bulk matter alone. We thus focus
on four possible regimes of the bulk reference and we check whether the
same classification still holds for the disordered arrangements where we
believe that multiple scattering between otherwise independent active nano
resonators will be the leading order effect in solid state random laser
samples of a high filling~fraction.

As a result we find that large scale systems can be driven with {\em subcritical}
excitation strengths up to the laser transition. Here we show the case for $P\,=\,0.2\,\gamma_{21}$
... $P\,=\,0.5\,\gamma_{21}$. The~pump power is definitely far below the
microscopic laser threshold and thus is {\it
  subcritical}. For~$P\,=\,0.2\,\gamma_{21}$ we obtain a symmetrically shaped
coherence volume with respect to the z-axis with a radius in the sample's
center, $Z\,=\,0$, of~$\xi\,=\,103.212\, \mu m$, see
Figure~\ref{vol}. Thus the coherence volume has an extent in the sample's center
of $2\cdot\xi\,=\,206.424 \, \mu m$. By~increasing the pump rate to
$P\,=\,0.5\,\gamma_{21}$ the coherence length reduces quantitatively to
$\xi\,=\,92.343\, \mu m$. At~the threshold value of the bulk material
$P\,=\,1.0\,\gamma_{21}$ we find qualitatively the same behavior; however,
the quantity is reduced again to $\xi\,=\,80.201\, \mu m$. By~increasing the pump
power to $P\,=\,0.5\,\gamma_{21}$ we find the coherence length of
$\xi\,=\,37.90\, \mu m$.  The~length scale $\xi$ is a direct measure of the
random laser mode in stationary state under specific conditions. It~is also a measure of the Cooperon
contribution, which is the perfect interference of a photon and its time
reversal in multiple scattering procedures that acts as the stimulation
process in a random laser. The~coherence volume is thus a measure of the {\em
  correlation} of multiple scattering events between otherwise independent
scatterers. At~the samples boundaries the coherence length $\xi$ becomes always
finite. The~coherence length $\xi$ is derived in the stationary state, which
means that it is the quantitative length scale which is the measure for the
system's random laser threshold. We find no crossover of these length scales
$\xi$ for varying pump rates anywhere in the~sample with otherwise identical parameters.

\subsection{Scattering mean-free Path and Diffusion Coefficient at the Random
  Laser~Threshold}
\label{sec:NUMDLS}

\begin{figure}[H]
\centering\hspace{0.8cm}\\  
\vspace*{0.5cm}
\centering{\scalebox{0.38}{\includegraphics{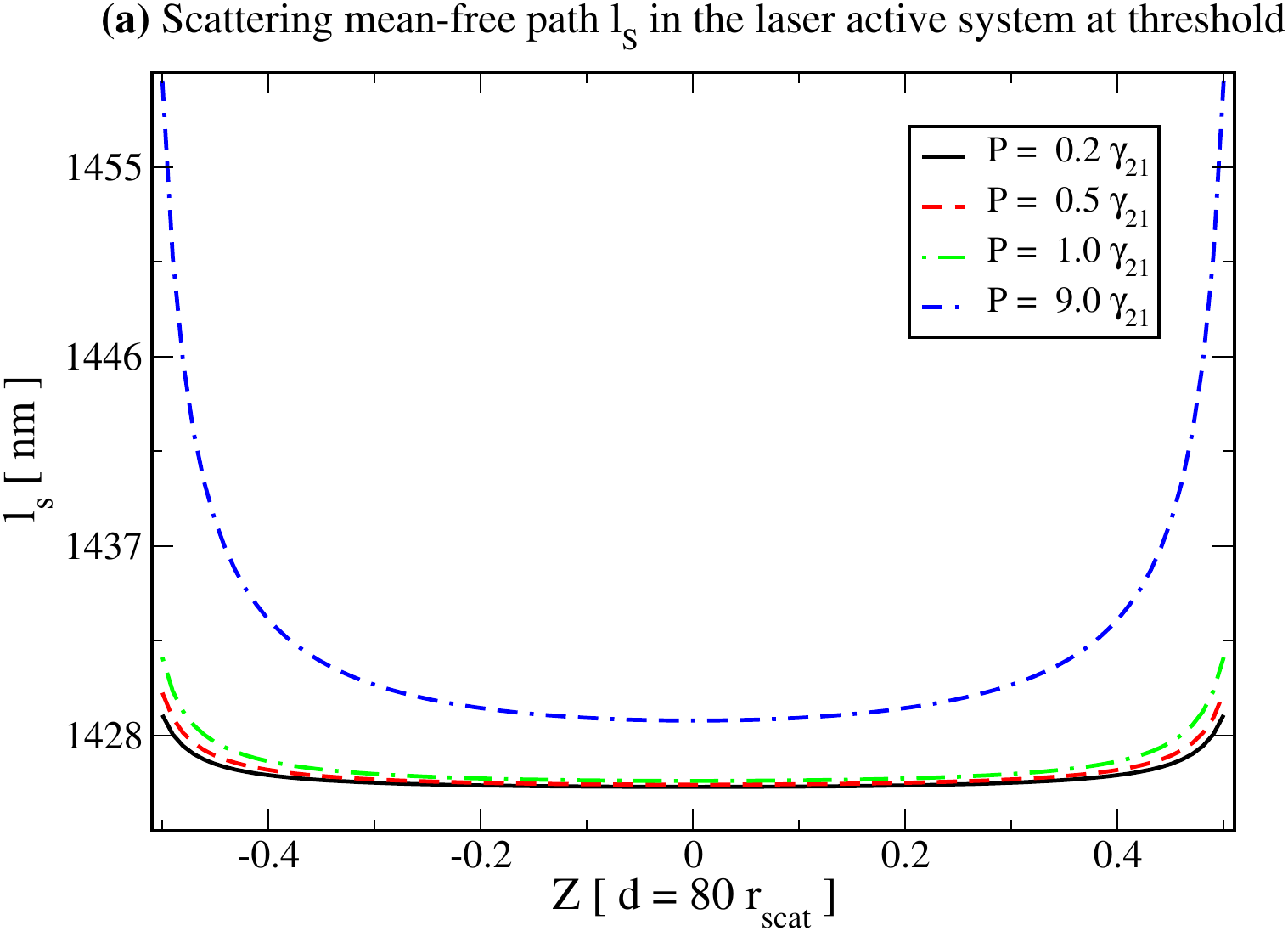}}}\\
\vspace*{0.5cm}
\centering{\scalebox{0.38}{\includegraphics{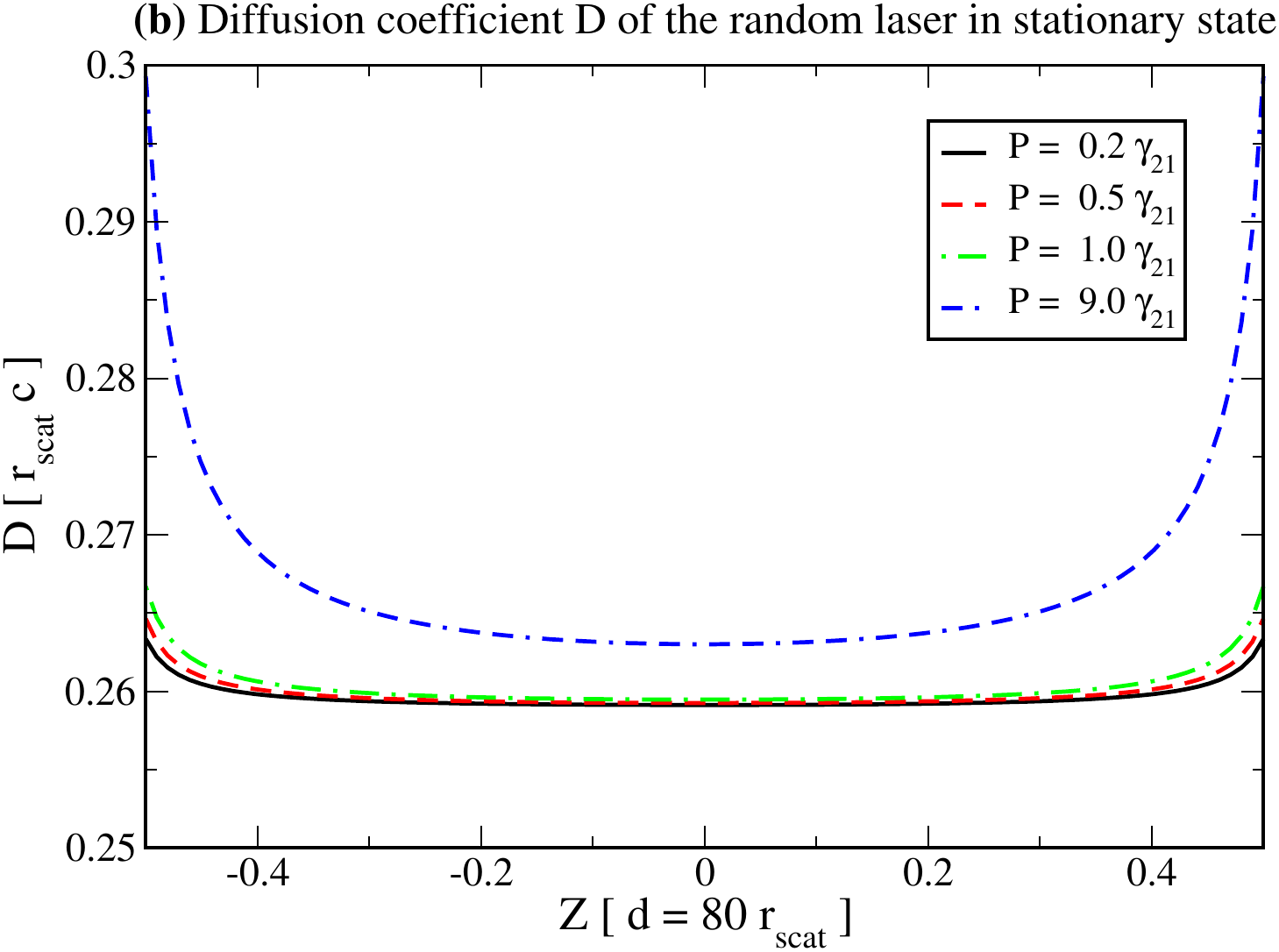}}}
\caption{(\textbf{a})  Scattering mean-free path $l_s$ in the laser active system
  at threshold. 
 In-depth dependent scattering mean-free path $l_s\,=\,\frac{1}{2{\rm
        Im}[{\sqrt{q^2+i{\rm Im}\Sigma(\omega)}}]}$, for~details see
~\cite{Frank11}, in~dependency to the excitation strength $P$. Parameters
    are identical to those of Figure~\ref{vol}. We find an in-depth dependent renormalization of $l_s$ 
    which is increased with 
    strength $P$. Both quantities, the~mean-free path $l_s$ and the diffusion
    coefficient $D(\Omega\,=\,0;Z)$, Figure~\ref{fig.3}b, are self-consistently derived and thus show
    quantitatively a different behavior at the sample boundaries $Z\,=\,\pm 0.5d$,
    rather than at $Z\,=\,0$. At~the open
    boundaries both characteristics $l_s$  and $D(\Omega\,=\,0;Z)$
    are increased. This behavior comes, however, along with the huge but inverse effect in the
    characteristics of the coherence volume, compare Figure~\ref{vol}.\\ (\textbf{b}) Diffusion coefficient $D$ of the random
  laser in stationary state.  
  In-depth dependent diffusion coefficient
  $D(\Omega\,=\,0;Z)$ as a material characteristic of the random laser
  arrangement at the laser threshold in~the stationary state. Parameters are
  identical to Figures~\ref{vol} and~\ref{fig.3}a. In~the excitation regime
  $P\,<\,1.0\,\gamma_{21}$ a moderate but existing dependency with respect to the external
  excitation and the position in-depth is found. In~the regime for excitation
  above the microscopic laser threshold $P\,<\,9.0\,\gamma_{21}$ we find an increasing
  deviation of the value of the diffusion coefficient D at the sample
  boundary of more than $7\,\%$ as compared to the transport wavelength. This
  deviation is only correlated to the absolute excitation power and to the
  open boundaries of the sample where $85\,\%$ of all reemitted photons
  shall be lost, whereas the in-depth sample of all photons multiplies
  scattering. Both results of $l_s$ and of $D$, which is finite, are signs of the
  absence of Anderson localization~\cite{Maret}.}\label{fig.3}
\end{figure}

We show in Figure~\ref{fig.3}a the result for the scattering mean-free path $l_s$
that is computed within the self-consistent framework in dependency to the
pump strength P and the position inside the sample. We find a significant
qualitative difference to the same length scale in the non-pumped regime. The~length scale $l_S$  at the random laser threshold, thus in stationary state,
is increased at the sample's lossy boundaries. This effect relies here
exclusively on the
incorporation of a loss rate at the boundaries, whereas the sample's filling is
$50\, \%$ volume filling everywhere in the slab. We find that the increase of
$l_s$ is qualitatively not directly inverse to the behavior of the coherence
or correlation length $\xi$; however it depends on the strength of the
excitation power. This is intuitively clear, since with the increasing
excitation power  of the
pump the refractive index of the single active Mie resonator is modulated. The~sample is homogeneously pumped, all parameters are equal to the case of
Figure~\ref{vol}.  We also find a significant quantitative
difference of the scattering mean-free path $l_s$ at the laser threshold,
which is for $P\,=\,9.0 \gamma_{21}$ at the boundaries, $l_s\,=\,1460.0$ nm and
in the center of the slab, at~$Z\,=\,0$, $l_s\,=\,1432.0$ nm. Thus $l_s$ under
stationary state lasing conditions is increased for our parameter set here by a factor of
$4.7$ to a factor of $4.87$ compared to the measure $l_s$ in the non-pumped
system, whereas its quality in terms of the derived formula is exactly the
same, compare reference~\cite{Frank11}. 

Our results for $l_s$,  Figure~\ref{fig.3}a, and~$D$,  Figure~\ref{fig.3}b, are as such remarkable, since they 
provide some insight on the physical consequence of the pumping of complex
active disordered media
with respect to their localization characteristics. The~measure for
localization is always the self-consistent diffusion coefficient $D$, that we
show as a result for the same parameters as for the calculation of $l_s$,
Figure~\ref{fig.3}a, and for the calculation of $\xi$,
Figure~\ref{vol}, in~ Figure~\ref{fig.3}b. The~diffusion coefficient shows qualitatively the same
behavior as $l_s$ in stationary state. It assumes quantitatively a finite value
which depends on the position and on the pump strength. We thus
conclude that random lasers at the stationary state undergo the lasing transition, which is, however, neither correlated nor equal to the
phase transition of Anderson~localization.

\subsection{Material-Dependency of the Mie Characteristics in Multiple
  Scattering Random~Lasers}
\label{sec:Cross}

We know from the results for the scattering mean-free path $l_s$ for photonic
transport in the weak excitation regime that a crossover of $l_s$ is expected
for various Mie resonators of a differing refractive index $n$ and thus a differing
permittivity $\epsilon_{scat}$. The~result for the calculation of the scattering mean-free path~\cite{Frank11} shows smooth but pronounced dips when the Mie resonance
condition for the single scatterer is approached and finally met. This Mie
resonance condition depends on the refractive index of the scatterer; its position
will thus shift in the energy spectrum for various materials and identical
scatterer parameters otherwise. Under~an ideal choice of parameters an almost
perfect point symmetry with respect to the crossover of the scattering mean
free path $l_s$ for two different refractive indices is~possible. 

Here we refer to the literature, references~\cite{Cao99,CAOPL}, where the
gain mechanism in solid state random lasing is vividly discussed with a focus
on two possible mechanisms, microscopic gain and gain due to artificial but
randomly 
formed laser cavities in the sense of a build up of correlated chains of
single scatterers. Whereas for the case of microscopic gain the properties of
the single Mie scatterer should change with respect to the resulting change of
the refractive index, in~the case of artificial but randomly
formed laser cavities the properties of the single scatterer should remain
rather~unchanged.

\begin{figure}[H]
\centering\hspace*{-1.0cm}{\scalebox{0.4}{\includegraphics{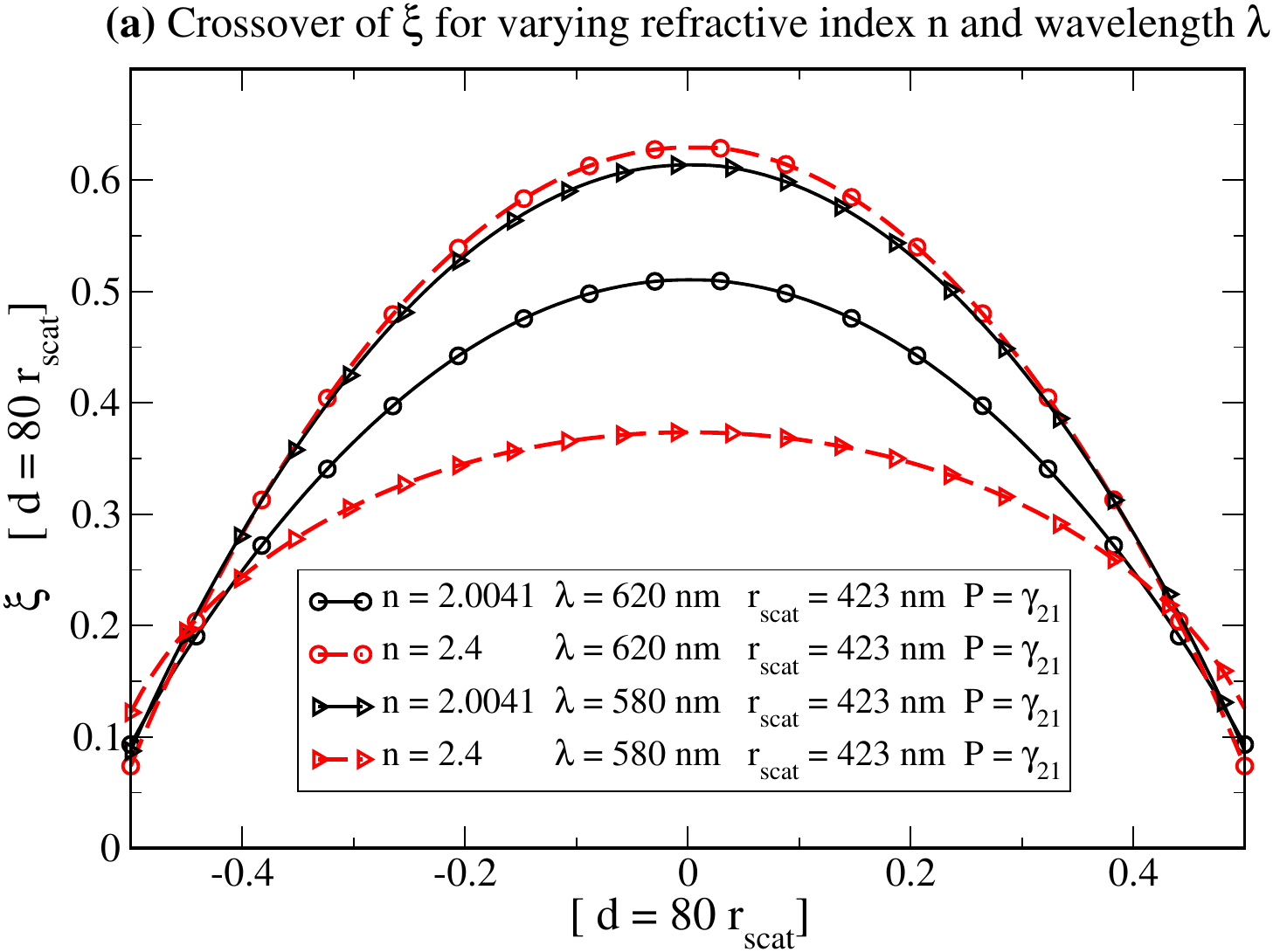}}}\\
\vspace*{0.5cm}
\centering\hspace*{0.0cm}{\scalebox{0.4}{\includegraphics{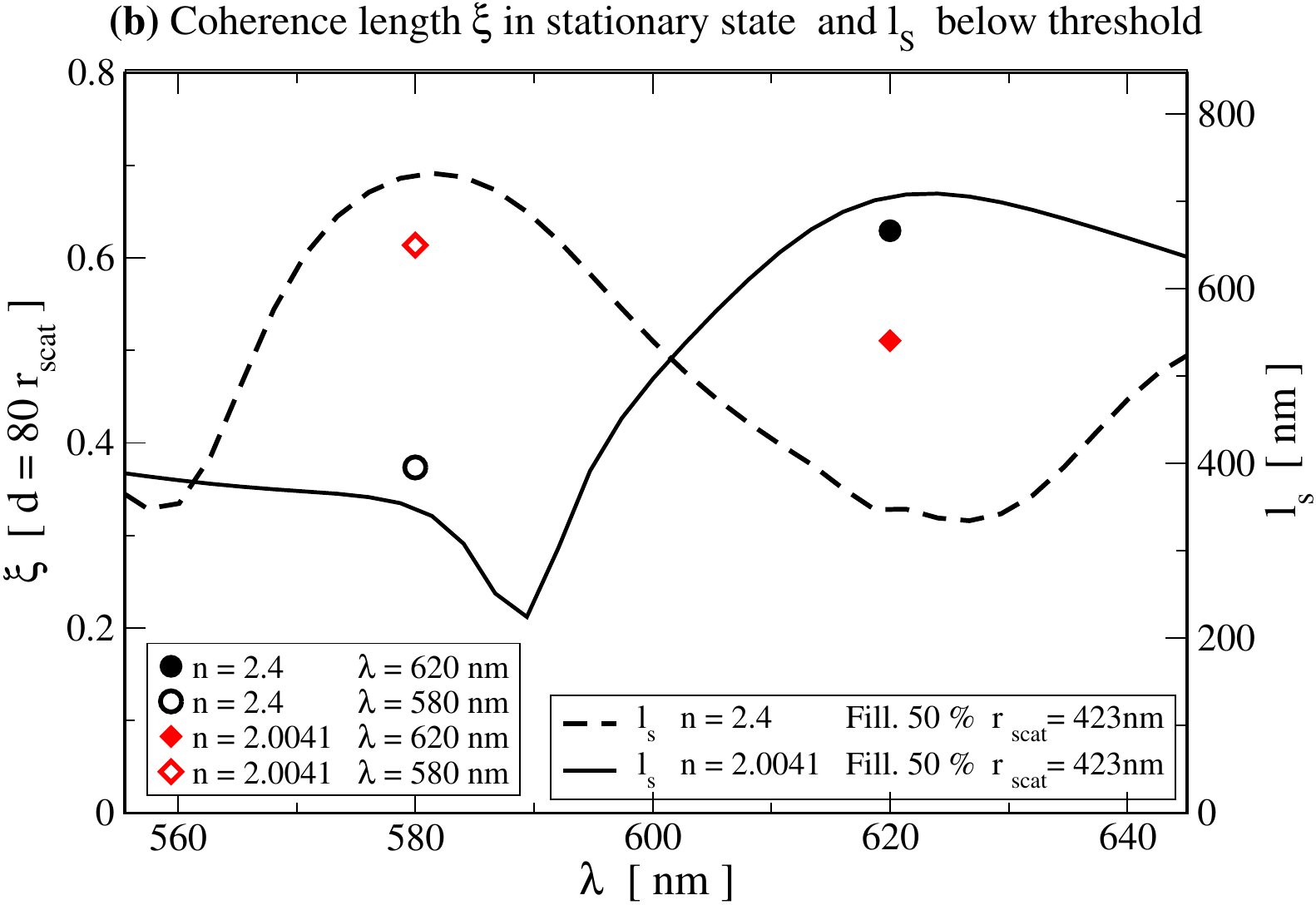}}}\\\vspace*{0.5cm}
\caption{(\textbf{a}) Crossover of the coherence length $\xi$ for
  various refractive indices. The crossover of the coherence length $\xi$ for two materials
  $n\,=\,2.4$ (C) and $n\,=\,2.0041$ (ZnO) is displayed for the transport
  wavelengths $\lambda\,=\,580.0$ nm and $\lambda\,=\,620.0$ nm in-depth
  dependent of the random laser sample. The~system parameters apart from the
  refractive index $n$ and the transport wavelength $\lambda$ are in either
  case identical, the~scatterers radius is $r_{scat}=\,423.0$ nm. The~  samples finite extent in z-direction is of $80\,r_{scat}$, the~excitation
  strength $P\,=\,1.0\,\gamma_{21}$ is
  moderate. A~principal crossover behavior of the length scale $\xi$
  is derived. We also find for the refractive index $n=\,2.4$ a
  crossover of the coherence lengths  $\xi$  near the sample's boundaries for $\lambda\,=\,580.0$ nm and
  $\lambda\,=\,620.0$ nm in the stationary state.\\ (\textbf{b}) Coherence length $\xi$ compared to the scattering mean-free
  path $l_s$ below~threshold. The crossover of the coherence length $\xi$ is
  shown in the samples
  center, $Z=0$, parameters are identical to
  Figure~\ref{fig.4}a. Intuitively a crossover behavior of $\xi$ is
  expected from the investigation of the scattering mean-free path $l_s$,
  $n\,=\,2.4$ (C) (dashed line), $n\,=\,2.0041$ (ZnO) (solid line),
  for disordered arrangements of Mie scatterers below the laser \mbox{threshold
~\cite{Frank06,Frank11}}. Opposed to our results for $l_s$ at the stationary
  state, see Figure~\ref{fig.3}a, we compare here to $l_s$  in the
  weakly pumped case. In~the spectral region between $\lambda\,=\,580.0$ nm and
  $\lambda\,=\,620.0$ nm we can confirm the crossover, which means that the
  Mie characteristics persist as a leading order effect under these specific
  conditions also while the system undergoes the lasing transition.}\label{fig.4}
\end{figure}

 This discussion for sure will be finally decided by novel
  experiments, which check the spatial extent of random laser modes in
  stationary state systematically by varying the geometric properties of the
  laser sample. So far we can deduce from our results that
for solid state random lasers gain
due to artificial but randomly formed laser cavities is not the leading order
gain mechanism, since~the refractive index is obviously changed
and the scattering mean-free path in stationary state is increased in
comparison to the same system under weak excitation, compare Section~\ref{sec:NUMDLS}. We check this behavior theoretically in order to
propose possible experiments which may be feasible to address the issue, and~one of them is to compare the measurable result of the coherence volume to the
crossover behavior of the scattering mean-free path for the same but only
weakly excited, so non-pumped, system in a
frequency range where we know the crossover of $l_s$ in the weakly excited
case is very pronounced. For~the system paremeters $r_{scat}\,=\,423.0$ nm,
the samples extent in z-direction of $d\,=\,80\,r_{scat}\,=\,33.84\,\mu m$, a~filling fraction of $50\,\%$ and the refractive indices of $n\,=\,2.4$  (C)
and  $n\,=\,2.0041$  (ZnO) such
 a very clear crossover of $l_s$ is found at $\lambda\,=\,600.0$ nm, compare
Figure~\ref{fig.4}b. We show in Figure~\ref{fig.4}a the in-depth dependency of the
correlation length $\xi$ for both systems, C and ZnO, under moderate pumping
$P\,=1.0\,\gamma_{21}$. The~pump wavelength as well as the transport wavelength
are  chosen as $\lambda\,=\,580.0$ nm and $\lambda\,=\,620.0$ nm. We find two qualitatively
different crossover situations. First~we find an obvious crossover of $\xi$
for the parameters $n\,=\,2.4$ and  $\lambda\,=\,580.0$ nm in the outer
selvedge of the sample, which is in the outer $10\,\%$ of the samples
extent on either open boundary. Second we find a qualitative change of  the coherence length
$\xi$ when we consider a different material by changing the passive refractive index from $n\,=\,2.0041$ to
$n\,=\,2.4$. For~the parameters $\lambda\,=\,580.0$ nm and $n\,=\,2.0041$ the
coherence length $\xi$ all over the sample depth obtains almost
the same characteristics and the same value as $\xi$ for the case of $n\,=\,2.4$ and
$\lambda\,=\,620.0$ nm. The~direct counterparts, $\xi$ for the parameter set of
$n\,=\,2.0041$, $\lambda\,=\,620.0$ nm, and~on the other hand $\xi$  for the set of $n\,=\,2.4$ and
$\lambda\,=\,580.0$ nm show a significant difference of about $40\,\%$. This
crossover becomes more obvious, when we
display the coherence lengths $\xi$, as~single marked points, in~comparison to the scattering mean-free
path $l_s$ for the weakly excited regime, marked as the solid and the dashed
line in Figure~\ref{fig.4}b. Whereas the absolute value of  $l_s$ is
rather symmetric for the refractive indices $n\,=\,2.0041$ and $n\,=\,2.4$ in the
spectral positions of $\lambda\,=\,580.0$~nm, and~of 
$\lambda\,=\,620.0$ nm, to~their crossover at $\lambda\,=\,600.0$ nm,  the~coherence length by contrast shows a crossover as well, but~the
almost perfect point symmetry behavior of the scattering mean-free path in the
weakly excited case is not present in the crossover behavior of $\xi$ in stationary
state. The~crossover
of  $\xi$ is also not confirmed for any crossover point of $l_s$ for weakly
excited Mie resonators in the~spectrum.

\subsection{Power Law Behavior of the Correlation Length Scale in Stationary~State}
\label{PowerLaw}

\begin{figure}[H]
\hspace*{0.99cm}\scalebox{0.45}{{\includegraphics[clip]{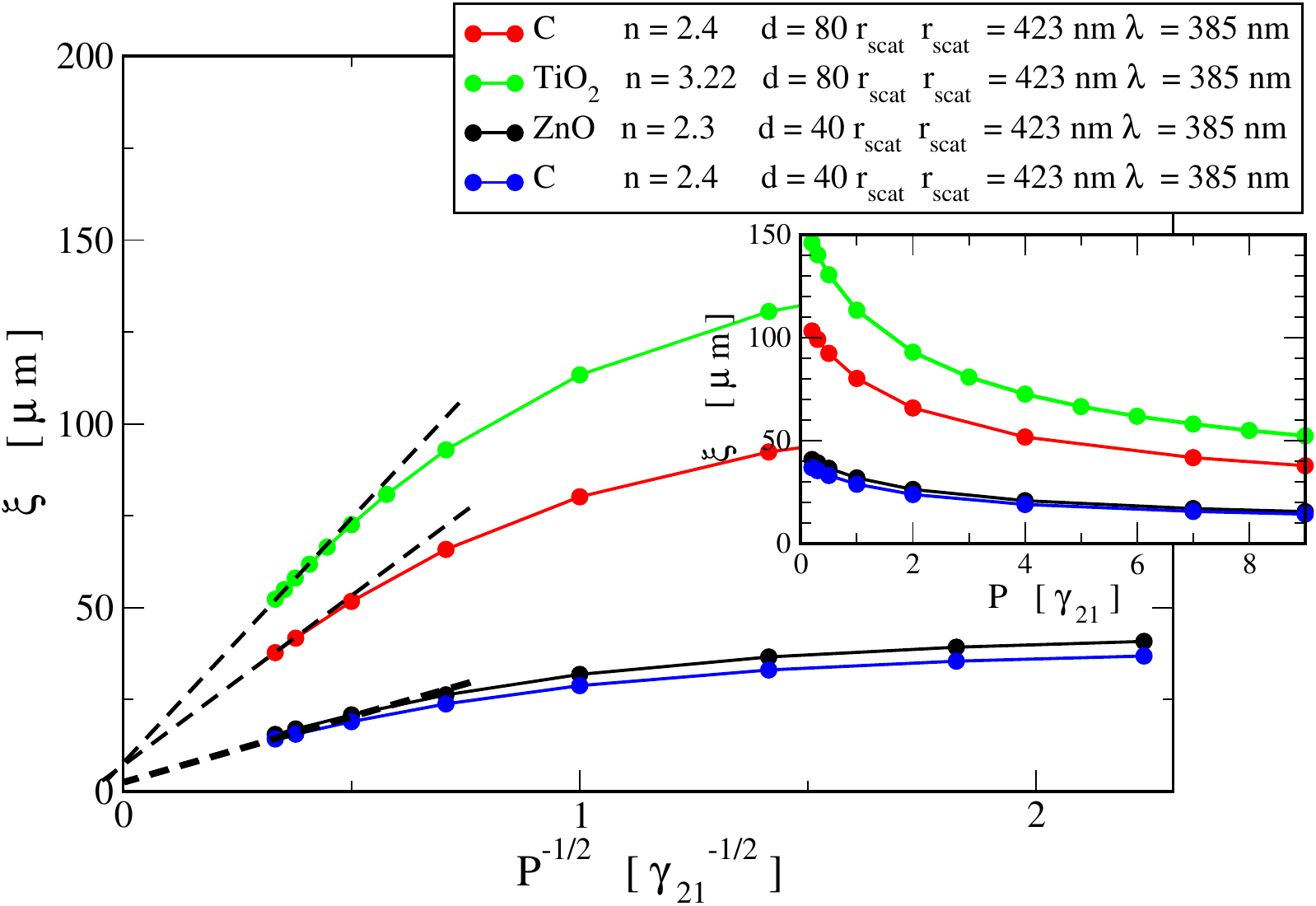}}}
\caption{Self-consistent coherence length $\xi$ of the random
  laser system in the finite size slab arrangement with a filling fraction of
  50 $\%$ for four different setups. We show $\xi$ at the samples
  center $Z\,=\,0$ for various extents of the slab with its dependency to the
  excitation power. Parameters are $d\,=\,40 r_{scat}$ and
  $d\,=\,80\,r_{scat}$, $r_{scat}=\,423.0$ nm.  Results are shown
for identical scatterers' radii of $ZnO$ (zinkoxide, $n=2.3$), of~$C$ (diamond,
$n=2.4$) and of $TiO_2$ (titania, $n=3.22$) active Mie scatterers without any
amplifying or absorbing background, $\epsilon_b\,=\,1.0$. With~increasing
excitation power $P$ a transition to a power law behavior of $\xi$ with
$1/\sqrt P$ is found in either of the setups (see black dashed lines). At~the
transport wavelength $\lambda=385.0$ nm we find that the Mie characteristic is a leading order effect in stationary state lasing. We find for the
scatterers radius $r_{scat}=\,423.0$ nm and $\lambda=385.0$~nm  for C, $n=2.4$, and~titania, $n=3.22$, a~significant difference in $\xi$ which cannot be explained
by bulk effects. We attribute it to enhanced multiple scattering for the
diamond Mie resonators with these parameters.}
\label{ZNOvsDiamont}
\end{figure}

We have investigated the coherence length $\xi$ with increasing pump power for
various refractive indices and various sample extents. In~Figure~\ref{ZNOvsDiamont} we show the coherence length  $\xi$ for the scatterers arrangements of $r_{scat}\,=\,423.0$ nm, filling fraction of $50\,\%$
and the samples extents of $d\,=\,40\,r_{scat}\,=\,16.92\,\mu m$ and
$d\,=\,80\,r_{scat}\,=\,33.84\,\mu m$. We show samples for $n\,=\,2.4$ (C),
$n\,=\,2.3$ (ZnO)  and $n\,=\,3.22$ (TiO$_2$). The~excitation wavelength is
$\lambda\,=\,385.0$ nm; results are displayed for pump powers from
$0.2\,\gamma_{21}$ up to $9.0\,\gamma_{21}$. We display the dependency of
$\xi$ with $1/\sqrt P$ and we find a characteristic power law behavior for
increasing pump power $P$. Another, similar power law has been found in experiments of
random lasers by Cao~et~al.~\cite{CAOPL}. When we compare several materials we find the specific
crossover behavior for the coherence length $\xi$ for the  samples of $n\,=\,2.4$ (C) and $n\,=\,3.22$ (TiO$_2$) whereas the power
law behavior is confirmed in either sample. The~coherence length $\xi$ for
TiO$_2$ assumes values of up to $\xi\,=\,150.0\,\mu m$; thus the coherence
volume has an extent in this case of up to $\xi\,=\,300.0\,\mu m$ in stationary
state.

\unskip

\section{Conclusions}

We have presented in this article a  quantum field theoretical approach for
three dimensional solid state random lasers comprised of active complex Mie
resonators. Thus we have implemented the Ward identity for non-conserving
media. The~systems are open at the boundaries. As~a result we
derived the coherence volume of random lasers as the
spatial characteristics of the random laser threshold in the stationary
state. This includes the spatial dependency of the self-consistent laser
gain. We~conclude from our results that the random laser of densely packed
strongly scattering random media is significantly depending on multiple
scattering processes in the sense that stationary state lasing can be reached
already for subcritical pump intensities, far below what is known as the microscopic
laser threshold of the bulk material. This characteristics is also depending
on Mie scattering characteristics of the active single scatterer. We have also presented the
results for the diffusion coefficient and the scattering mean-free path
which have been derived by the self-consistent framework for finite samples in
the stationary lasing state and which are thus
also spatially dependent. We find a significant deviation of the scattering
mean-free path in the lasing regime in stationary state from $l_s$ the weakly
excited case. This deviation is confirmed in the characteristics of the
diffusion coefficient which is finite and spatially dependent. We~compared
the qualitative behavior of the coherence length of the stationary state
random laser to the qualitative behavior of the scattering mean-free path of
photons in the same but non-pumped and thus non-inverted setup. We find that the
  coherence length in stationary state qualitatively follows the material
  dependent spectral crossover
  behavior of the scattering mean-free path of disordered samples of weakly
  excited Mie resonators and otherwise identical parameters in the case of
  very pronounced Mie characteristics in weakly excited systems. We also compared the coherence length in stationary state
for identical scatterers of various materials, diamond (C), zinkoxide
(ZnO) and titania (TiO$_2$), for~increasing pump strength as well as for varying
samples thicknesses. We find in
all our results a power law behavior of the coherence length with respect to
the pump intensity. We conclude from our theory that for the
  understanding of the physics of random
  lasers of multiple scattering ensembles, the concept of the localization
  length which is a fundamental characteristics for
mesoscopic transport of light in random media is conceptually not
sufficient since it does not provide knowledge about the effect of self-consistent gain including the microscopic electronic subsystem. The~length scale of the coherence or
correlation length and the coherence or correlation volume, respectively, are 
more comprehensive scales for laser active random media and they characterize
the laser modes and the laser threshold
behavior. We investigated in
this work arrangements of independent monodisperse Mie resonators. It will be the
subject of future work to investigate the influence of gain and absorption
with respect to sub- and hyper-diffusion systems and with respect to laser
samples of quasi-ordered clusters and {\it meta glasses}.

%
%

\vspace{6pt} 


\vspace{+6pt}

\authorcontributions{All authors contributed equally to this work. All authors
wrote and reviewed the~manuscript.}
\funding{This research received no external funding.}

\acknowledgments{We thank H. Cao,  B. A. van
  Tiggelen, K. Busch and J. Kroha for very fruitful~discussions.}

\conflictsofinterest{No conflicts of interest.} 



\reftitle{References}





\end{document}